\documentclass[prl,twocolumn,showpacs,superscriptaddress]{revtex4}

\usepackage{graphicx}
\begin{document}

\title{Critical and non-critical jamming of frictional grains}

\author{Ell\'ak Somfai}\thanks{Present address: Department of Physics, Oxford University,  1 Keble Roard, Oxford, OX1 3NP
 United Kingdom} \affiliation{Instituut--Lorentz,
Universiteit Leiden, Postbus 9506, 2300 RA Leiden, The Netherlands}

\author{Martin van Hecke} \affiliation{Kamerlingh Onnes Lab,
Universiteit Leiden, Postbus 9504, 2300 RA Leiden, The Netherlands}

\author{Wouter G. Ellenbroek} \affiliation{Instituut--Lorentz, Universiteit
Leiden, Postbus 9506, 2300 RA Leiden, The Netherlands}

\author{Kostya Shundyak} \affiliation{Instituut--Lorentz, Universiteit
Leiden, Postbus 9506, 2300 RA Leiden, The Netherlands}

\author{Wim van Saarloos} \affiliation{Instituut--Lorentz,
Universiteit Leiden, Postbus 9506, 2300 RA Leiden, The Netherlands}

\date{\today}

\begin{abstract}
We probe the nature of the jamming transition of frictional
granular media by studying their vibrational properties as a
function of the applied pressure $p$ and friction coefficient
$\mu$. The density of vibrational states exhibits a crossover from
a plateau at frequencies $\omega \gtrsim \omega^*(p,\mu)$ to a
linear growth for $\omega \lesssim \omega^*(p,\mu)$. We show that
$\omega^*$ is proportional to $\Delta z$, the excess number of
contacts per grains relative to the minimally allowed, isostatic
value. For zero and infinitely large friction,
typical packings at the jamming threshold  have $\Delta z \rightarrow 0$,
and then exhibit critical scaling.
We study the nature of the soft  modes in these two limits, and find
that the ratio of elastic moduli is governed by
the distance from isostaticity.
\end{abstract}

\pacs{ 45.70.-n, 
46.65.+g, 
83.80.Fg 
}

\maketitle

Granular media, such as sand, are conglomerates of dissipative,
athermal particles that interact through repulsive and frictional
contact forces. When no external energy is supplied, these
materials jam into a disordered configuration under the action of
even a small confining pressure \cite{jamming_nature}. In recent
years, much new insight has been amassed for the jamming
transition of models of deformable, spherical, athermal, {\em
frictionless} particles in the absence of gravity and shear
\cite{epitome12}. The beauty of such systems is that they allow
for a precise study of the jamming transition that occurs when the
pressure $p$ approaches zero (or, geometrically, when the particle
deformations vanish). At this jamming point ``J'' and for large
systems, the contact number \cite{footnote} equals
the so-called isostatic value $z^0_{\rm iso}$ (see below), while
the packing density $\phi^{0}_{\rm J}$ equals random close packing
\cite{epitome12,moukarzelwitten}. Moreover, for compressed systems
away from the jamming point, the pressure $p$, the excess contact
number $\Delta z=z(p) -z^{0}_{\rm iso}$ and the excess density
$\Delta \phi = \phi-\phi^{0}_{\rm J}$ are related by  powerlaw
scaling relations --- any of the parameters $p,\Delta z$ and
$\Delta \phi$ is sufficient to characterize the distance to
jamming.

Isostatic solids are marginal solids
--- as soon as contacts are broken, extended ``floppy
modes'' come into play \cite{alexander}. Approaching this marginal
limit in frictionless packings as $p\rightarrow 0$, the density of
vibrational states (DOS) at low frequencies is strongly enhanced
--- the DOS has been shown to become essentially constant up to
some low-frequency crossover scale $\omega^*$, below which the
continuum
scaling $\sim \omega^{d-1}$
is recovered
\cite{barrat,epitome12,wyartetc,silbert,leo_matthieu,ellak,wyart}. For
small pressures, $\omega^*$ vanishes $\sim \Delta z$. This signals
the occurrence of a critical lengthscale, when translated into a
length via the speed of sound, below which the material deviates
from a bulk solid \cite{leo_matthieu}. The jamming transition for
frictionless packings thus resembles a critical transition.

In this paper we address the question whether an analogous
critical scenario occurs near the jamming transition at $p\!=\!0$
of {\em frictional} packings. The Coulomb friction law states that
when two grains are pressed together with a normal force $F^{\rm
n}$, the contact can support any tangential friction force $F^{\rm
t}$ with $F^{\rm t} \leq \mu F^{\rm n}$, where $\mu$ is the
friction coefficient.
In typical packings, essentially none of these tangential forces
is at the Coulomb threshold $F^{\rm t}=\mu F^{\rm n}$
\cite{silbert,Kasahara}.
A crucial feature of these packings of frictional
particles is that, for $p\rightarrow  0$,  they span a range of
packing densities and have a nonunique contact number
$z_{\rm J}(\mu)$, which typically is larger than 
the frictional isostatic value $z^{\mu}_{\rm
iso}=d+1$ \cite{unger,makse,rlp,Kasahara} (see below).
So two questions arise: do frictional systems ever
experience a ``critical'' jamming transition in the sense that
$\omega^*$ vanishes when $p\rightarrow 0$? What is the nature of the
relations between $p$, $\mu$, $\omega^*$, excess contact number
$z(\mu,p)-z_{\rm J}(\mu)$ and excess density? 

\begin{figure*}[tb]
\includegraphics[width=17.6cm]{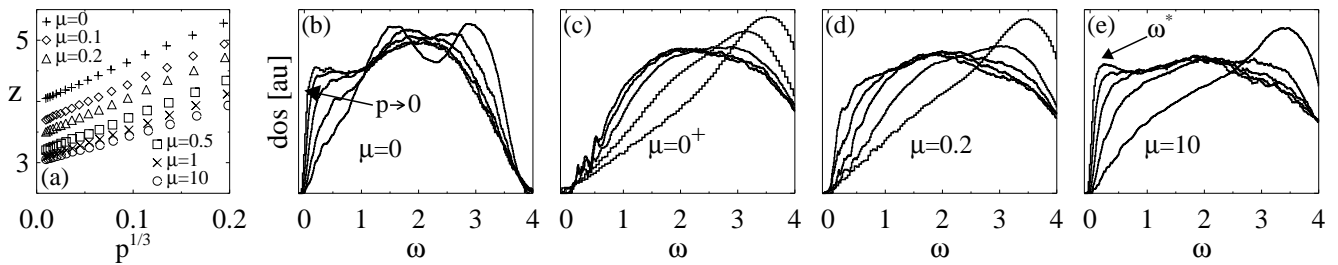}
\vspace*{-0.5cm}

 \caption{(a) Average contact number $z$ as a function of $p^{1/3}$
for various $\mu$ as indicated. (b-e) Vibrational DOS for granular
packings for friction coefficients as indicated, and for pressures
approximately
$5\times10^{-6},5\times10^{-5},5\times10^{-4},4\times10^{-3}$ and
$3\times10^{-2}$. For decreasing $p$ the DOS becomes steeper for
small $\omega$, and the crossover frequency $\omega^{*}$, indicated
in (e), decreases with $p$. The packing with $\mu = 0^+$ is obtained
by first making a frictionless packing and then turning on the
tangential frictional forces in the DOS calculation. As noted in the text, all
frequencies are scaled by a factor $p^{1/6}$.} \label{fig1}
\end{figure*}

The results which we present below give convincing evidence that
jamming of frictional grains should be seen as a two-step process.
The first step is the selection of $z$
for fixed $\mu$ and a given numerical procedure (see Fig.~1a);
this has been studied before \cite{unger,makse,rlp,Kasahara}.
Our focus here is on the second step, the fact that the critical
frequency $\omega^*$ of the DOS of vibrations of infinitesimal amplitude is proportional to the
distance to the frictional isostatic point $\Delta z:=z(\mu,p)-z_{\rm
iso}^{\mu}$. The crucial point is that $z_{\rm J}(\mu)$ and
$z_{\rm iso}^{\mu}$ in general  differ. In particular for small
values of $\mu$, the contact number saturates at a value
substantially above the isostatic limit, $\omega^*$ saturates at a
finite value and the system remains far from criticality. For
increasing values of $\mu$, however, $z_{\rm J}(\mu)$ approaches
$z_{\rm iso}^{\mu}$, and thus for large friction values,
$\omega^*$ evidences an increasingly large scale near the jamming
point.
The vanishing of $\omega^*$ has its origin in the emergence of
floppy modes at the isostatic point. 
We show that similarly to frictionless systems, both $\omega^*$ and the ratio
of shear to compression modulus $G/K$ scale as $\Delta z$. 
In short, the distance to isostaticity, which is well-defined, governs the
scaling of
both frictional and frictionless systems, providing a unified picture of jamming
of weakly compressible particles.

Let us, before presenting our results, recapitulate the well-known
counting arguments for the contact number in the limit
$p\rightarrow 0$ for dimension $d$ \cite{moukarzelwitten}. Since
the deformation of the spheres vanishes in the limit $p\!=\!0$,
all particles in contact are at a prescribed distance, which gives
$zN/2$ constraints on the $dN$ particle coordinates, leading to $z
\leq 2d$.  For the {\em frictionless} case, the $zN/2$ normal
contact forces are constrained by $Nd$ force balance equations ---
hence only for $z \geq 2d$ can we generically find a set of
balancing forces \cite{noterattlers}.
Taken together, this yields $z \rightarrow
2d=:z^0_{\rm iso}$ as $p\rightarrow 0$: at the jamming transition,
packings of frictionless spheres are isostatic. For {\em
frictional} packings, there are $zdN/2$ contact force components
constrained by $dN$ force and $d(d-1)N/2$ torque balance equations
--- thus $z \geq d+1$, with $z^\mu_{\rm iso}=d+1$ the isostatic
value. Hence, at the jamming transition, frictional spheres do not
have to become isostatic but can attain contact numbers between
$z^{\mu}_{\rm iso}=d+1$ and $2d$. While it is not well understood
what selects the contact number of a frictional
packing at J, simulations for discs in 2 dimensions 
show that in practice $z_{\rm J}(\mu)$ is a decreasing function
of $\mu$, ranging from 4 at small $\mu$ to 3 for  large $\mu$
\cite{unger,makse,Kasahara}; see also Fig.~\ref{fig1}a.

{\em Procedure ---} Our numerical systems are 2D packings of 1000
polydisperse spheres that interact through 3d Hertz-Mindlin forces
\cite{HMfoot}, contained in square boxes with periodic boundary
conditions. We set the Young modulus of the spheres $E^*=1$,
which becomes the pressure unit, and set the Poisson ratio to
zero. Our unit of length is the average grain diameter, the unit
of mass is set by asserting that the grain material has unit
density and the unit of time follows from the speed of sound of
pressure waves inside the grains \cite{ellak}. The packings are
constructed by cooling while slowly inflating the particle radii
in the presence of a linear damping force, until the required
pressure is obtained.  For each value of
$\mu$ and $p$, 20 realizations are constructed (occasional runs
with 100 realizations did not improve accuracy).

Once a packing is made, the additional damping force is switched
off and the dynamical matrix is obtained by linearizing the
equations for small amplitude motions, which include both
rotations and translations. It is
important to realize the special role of the friction: if
the density of contacts that precisely satisfy $F^{\rm
t}\!=\! \mu F^{\rm n}$ is negligible, the Coulomb condition
$F^{\rm t}\!\leq\! \mu F^{\rm n}$ only plays a crucial role during
the preparation of a packing. We will assume that this
is the case, and come back to this subtle point later. Under these
assumptions, and for arbitrarily small amplitude vibrations, the
Coulomb condition is automatically obeyed and the value of $\mu$
does not play a role anymore in analyzing the vibrational modes.
Moreover, the changes in $F^{\rm t}$ are then non-dissipative and
the eigenmodes of the dynamical matrix are undamped. In this
picture, the main role of the value of the friction
coefficient is in tuning the contact number.

We analyze the density of vibrational states (DOS) of the packings
thus obtained. Since for  Hertzian forces the effective spring
constants scale with the overlap $\delta$ as ${\rm d}F^{\rm n}/{\rm
d}\delta\! \sim\! \delta^{1/2}\!\sim\! p^{1/3}$ \cite{HMfoot}, all
frequencies will have a trivial $p^{1/6}$ dependence. To facilitate
comparison with data on frictionless spheres with one-sided harmonic
springs \cite{epitome12,silbert}, we report our results in terms of
scaled frequencies in which this $p^{1/6}$ dependence has been taken
out.

{\em Variation of $z$ ---} Anticipating the crucial role of the
contact number, we start by presenting $z(\mu,p)$ for our packings.
Fig.~1a confirms the earlier
observations \cite{unger,makse,Kasahara} that the effective value
of $z_{\rm J}(\mu) =:z(\mu,p\to 0 )$ varies from about 4 to about 3 when $\mu$ is
increased. Moreover, the excess number of
contacts $z(\mu,p)-z_{\rm J}(\mu)$ varies with pressure as $p^{1/3}$
for all values of $\mu$.

{\em DOS ---} Figs.~\ref{fig1}b-e show our results for the DOS for
various values of $\mu$. For the frictionless case shown in
Fig.~\ref{fig1}(b), we recover the gradual development of a
plateau in the density of states as the pressure is decreased
\cite{epitome12,silbert}. For this case $z\rightarrow z^{0}_{\rm
iso}$, and the crossover frequency $\omega^*$ scales as $\Delta
z=z-z^{0}_{\rm iso}$ \cite{wyartetc,silbert,leo_matthieu} (see
below). However, as Figs.~\ref{fig1}(c-d) illustrate, as soon as
the tangential frictional forces are turned on, this enhancement
of the DOS at low frequencies largely disappears, because the
frictionless ``floppy modes'' are killed. This point is
demonstrated most dramatically in Fig.~\ref{fig1}(c), where the
underlying packing has been generated for zero friction, and the
friction is only ``switched on'' when calculating the DOS
--- this represents the limit of vanishing small but nonzero
friction, for which the DOS is seen to be very far from critical. By
increasing the friction coefficient, the development of a plateau
and the scaling of the crossover progressively reappear
(Fig.~\ref{fig1}(e)). The intuitive picture that emerges is that
with increasing  friction, granulates at the jamming point approach
criticality.

\begin{figure}[tb]
\includegraphics[width=7.cm]{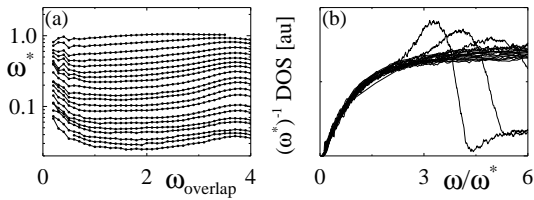}
\vspace*{-0.2cm}

 \caption{(a) $\omega^*$ for $\mu=10$ as function of $\omega_{\rm
overlap}$ --- the regime ($1<\omega_{\rm overlap}<3$) corresponds to
the crossover regime in the DOS that we focus on here (see text and
panel (b)). (b) The rescaled DOS for $\mu\!=\!10$ exhibit good data
collapse in the crossover regime. Here 20 rescaled DOS are shown
with $p$ ranging from 9$\times 10^{-7}$ to 3$\times 10^{-2}$.
}\label{fig2}
\end{figure}

In order to back this up quantitatively, we perform a scaling
analysis of the low-frequency behavior of these DOS. To avoid
binning problems, we work with the integrated density of states
$I(\omega)= \int^\omega {\rm d}\omega' DOS(\omega')$. The critical
frequencies are  then obtained by requiring that the rescaled
integrated DOS, $(\omega^*)^{-1}I(\omega/\omega^*)$, collapse. Such
collapse is never perfect, in particular since not all DOS have
precisely the same ``shape'' (Fig.~1). We  vary the value of
$\omega_{\rm overlap}:=\omega/\omega^*$ where we require the
rescaled integrated DOS to overlap --- as Fig.~\ref{fig2}a
illustrates, this yields precise values for $\omega^*$ as function
of $\omega_{\rm overlap}$. Restricting ourselves to  the crossover
regime ($1<\omega_{\rm overlap}<3$), we obtain by this procedure
both an estimate of $\omega^*$ and of its errorbar. As
Fig.~\ref{fig2}b illustrates, when rescaled with these estimated
values of $\omega^*$, the collapse of the DOS in the crossover
regime is convincing.

\begin{figure}[tb]
\includegraphics[width=8.5cm]{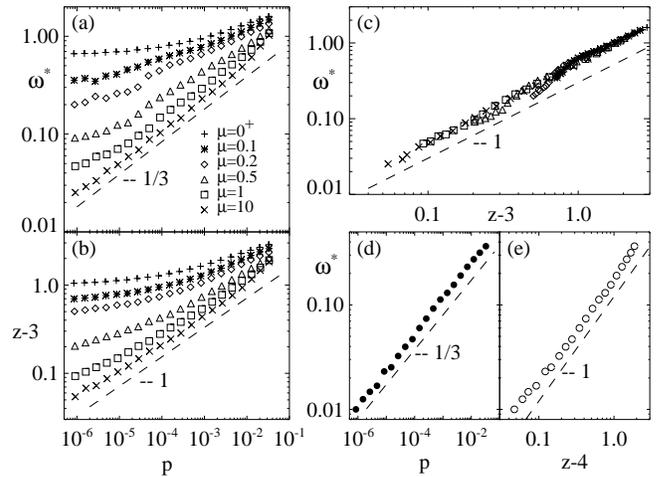}
\vspace*{-0.2cm}

 \caption{(a) $\omega^{*}$ as function of pressure $p$ for a range of
friction coefficients $\mu$ --- errorbars are similar or smaller
than symbol sizes (b) Deviation from isostaticity for the same range
of parameters. (c) $\omega^{*}$ scales linearly with the distance to
isostaticity for frictional packings. (d-e) $\omega^*$ for
frictionless packings scales both with $p$ and $z-4$. Dashed lines
indicate power laws with exponents as indicated. For details, see
text. }\label{fig3}
\end{figure}

{\em Scaling of $\omega^*$ ---} The first main result of this
paper is shown in Fig.~\ref{fig3}:  $\omega^*$ does {\em not}
scale in a simple way with $p$, but {\em the data for all $\mu$ and $p$
collapse onto a single curve when plotted as function of} $ \Delta
z = z-z^{\mu}_{\rm iso}$ (Fig.~\ref{fig3}a,c). Moreover, $\omega^*
\sim \Delta z$ --- the plot of $z-3$ versus $p$ shown in
Fig.~\ref{fig3}b is essentially equivalent to the plot of
$\omega^*$ vs $p$.  In other words, packings with $\Delta z\ll 1$
have many low-lying vibration modes and correspondingly a large
enhancement (plateau) in the DOS.  The dominant quantity governing
the behavior of frictional granular media is the distance from the
frictional ``critical point'' $z=z_{\rm iso}^{\mu}=d+1=3$. This
distance can be characterized most conveniently by $\Delta z$ ---
put in these
terms, the scaling of $\omega^*$ for frictional media is very
similar to the scaling for frictionless media shown in
Fig.~\ref{fig3}(d,e).

{\em Scaling of Elastic Moduli ---} The contact number for
isostatic systems reaches the minimum needed to remain stable ---
hence additional broken bonds then generate global zero-energy
displacement modes, so-called floppy modes
\cite{epitome12,alexander}. For frictionless systems, the excess
of soft modes and development of a plateau in the DOS for small
$\Delta z$ are intimately connected to these floppy modes
\cite{wyartetc,leo_matthieu,palpha}. 
For frictionless systems they also
cause the shear modulus $G$ to become much smaller than the bulk modulus $K$
--- in fact, $G/K \sim \Delta z$ \cite{epitome12,wyart,palpha}.

Our second main result is that we have found
numerically that for frictional systems the ratio $G/K$ only depends on
$\Delta z$ (and not on e.g., $\mu$), and for small $\Delta z$ it 
also scales as $\Delta z$. To calculate the moduli, we start from the
dynamical matrix which relates
forces and displacements. Calculating, in linear order, the global stress 
resulting from an imposed deformation, the bulk and shear moduli are 
deduced \cite{palpha}. Since for Hertzian forces the effective spring
constants scale as $p^{1/3}$, we have divided out this trivial
pressure dependence. The results of our calculations are shown in
Fig.~4. As could be expected, the (rescaled) bulk modulus $K$
remains essentially constant. Surprisingly, the shear modulus $G$
becomes much smaller than $K$ for small $p$ and large $\mu$, and
when plotted as function of $\Delta z$, the ratio $G/K$ is found
to scale as $\Delta z$,
as was predicted in \cite{wyart}.
Hence, in packings of deformable spheres,
both $\omega^*$ and $G/K$ scale with $\Delta z$, regardless of the
presence of friction.

{\em Discussion ---} The sudden change in the DOS when increasing
$\mu$ from zero hints at the singular nature of the $\mu \rightarrow
0$ limit. On the one hand, the nature of the dynamical matrix
suddenly changes in this limit because the rotational degrees of
freedom which are irrelevant for $\mu\!=\!0$ turn on as soon as $\mu
\neq 0$. On the other hand, we have recently found that the slower the
packings are allowed to equilibrate during their preparation, the more
the density of fully mobilized contacts, i.e., those for which $F^{\rm
t}\!=\! \mu F^{\rm n}$, tends to increase; in fact especially for
small $\mu$ the fraction of fully mobilized contacts in slowly
equilibrated samples becomes very substantial
\cite{kostya,footnotefullymob}.  The effect of these fully mobilized
contacts on the DOS depends on additional physical assumptions. For
example, if we assume that, for some reason, the contacts remain
constrained at the Coulomb threshold in the
vibrational dynamics, we expect them to have an enhanced DOS for small
pressures at all $\mu$. More likely, these contacts
would slip, leading to an initial strongly nonlinear response after
which no contacts would be fully mobilized anymore, and our results
for the DOS would go through essentially unchanged.

Note that even a small difference between dynamic and static friction
could suppress the effect of the fully mobilized contacts. Moreover, for
realistic values of the friction ($\mu \gtrsim 0.7$, say) these effects
are not very important since there the fraction of fully mobilized
contacts is small. Thus, the results of this paper will apply directly
to packings with experimentally relevant values of the friction.

A second issue that deserves further attention is the nature of
the soft modes. Our scaling result for $G/K$ suggest
that for frictional systems these are dominated by shear-like
(volume conserving)
deformations, just as for frictionless systems. 
Apparently, rotations and particle motions
couple such as to allow large scale floppy-mode-like distortions of frictional
isostatic packings. Indeed, numerically obtained low frequency
eigenmodes of frictional and frictionless systems look remarkably
similar. Whether the scaling of $\omega^*$, the scaling of $G/K$ and the
nature of floppy modes are similarly related in more general
systems, such as packings of frictionless or frictional
non-spherical particles, is an important question.

\begin{figure}[tb]
\includegraphics[width=7.4cm]{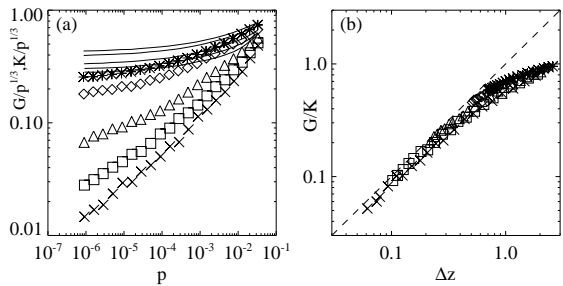}
\vspace*{-0.3cm}

\caption{Scaling of bulk modulus $K$ and shear modulus $G$ as
function of $p$ and $\mu$ --- similarly to the data for
$\omega^*$, the trivial $p^{1/3}$ dependence has been divided out.
(a) The rescaled bulk modulus $K$ (curve) essentially levels off
for small $p$, while the shear modulus $G$ (symbols as in Fig.~3)
varies strongly with both $p$ and $\mu$. (b) $G/K$ scales like the
excess contact number for small $\Delta z$.} \label{fig4}
\end{figure}

\enlargethispage{1cm}
{\em Outlook ---} Our study of the density of vibrational states
for frictional systems gives strong evidence for a scenario partly
analogous to the one for frictionless packings: frictional
granular media become critical and exhibit scaling when their
contact number approaches the isostatic limit.
But there is an important difference to the frictionless case:
while there the isostatic limit is automatically reached in the
hard particle/small $p$ limit, this is not necessarily so for the frictional
case --- here $p$ and $z$ are not directly related, and only for
large friction does $z$ approach isostaticity at small pressures.
This isostatic point is relevant in practice: most materials have
a value of $\mu$ of order 1, and as Fig.~\ref{fig3} and 4
illustrate, one then observes approximate scaling over quite some
range.

We are grateful to M. Depken, L. Silbert, S. Nagel, D. Frenkel, H. van der Vorst and T. Witten
for illuminating discussions. ES acknowledges support from the EU
network PHYNECS, WE support from the physics foundation FOM, and
MvH support from NWO/VIDI.



\begin{thebibliography}{99}

\bibitem{jamming_nature} A. J. Liu and S. Nagel, Nature {\bf 396}, 21 (1998).

\bibitem{epitome12} C. S. O'Hern {\em et al.},
Phys. Rev. E {\bf 68}, 011306 (2003); 
C. S. O'Hern {\em et al.},
Phys. Rev. Lett. {\bf 88},
075507 (2002).

\bibitem{footnote} We will use the convention that the upper index distuinguish
frictionless $(0)$ from frictional $(\mu)$ quantities, while lower
index indicates whether quantities are taken at the jamming (J) or
isostatic (iso) point.

\bibitem{moukarzelwitten} C. F. Moukarzel, Phys. Rev. Lett. {\bf 81}, 1634
(1998).

\bibitem{alexander} S. Alexander, Phys. Rep. {\bf 296}, 65 (1998).

\bibitem{barrat} A. Tanguy {\em et al.}, 
Phys. Rev. B {\bf 66}, 174205 (2002).

\bibitem{wyartetc} M. Wyart, S. R. Nagel and T. A. Witten,
Europhys Lett {\bf 72}, 486 (2005).

\bibitem{silbert} L. E. Silbert, A. J. Liu and S. R. Nagel,
Phys. Rev. Lett. {\bf 95}, 098301 (2005); L. E. Silbert, 
{\em et al.}
Phys.  Rev. E {\bf 65},
031304 (2002).

\bibitem{leo_matthieu} M. Wyart {\em et al.}
Phys. Rev. E {\bf 72}, 051306 (2005).

\bibitem{ellak} E. Somfai {\em et al.},
Phys. Rev. E {\bf 72}, 021301 (2005).

\bibitem{wyart} M. Wyart, Ann. Phys. Fr. {\bf 30}, 1-96 (2005).

\bibitem{Kasahara} A. Kasahara and
H. Nakanishi, Phys. Rev. E {\bf 70}, 051309 (2004).

\bibitem{unger} T. Unger, J. Kert\'esz, and D. E. Wolf, Phys. Rev. Lett. {\bf 94}, 178001 (2005).

\bibitem{makse} H. P. Zhang and H. A. Makse, Phys. Rev. E {\bf 72}, 011301 (2005);
H. A. Makse {\em et al.},
Phys. Rev. E {\bf 70}, 061302 (2004).

\bibitem{rlp} G. Y. Onoda and E. G. Liniger, Phys. Rev. Lett. {\bf
64}, 2727 (1990).

\bibitem{noterattlers} In the counting argument, $N$ refers to the
number of non-rattling particles.

\bibitem{HMfoot}
i.e., normal force $F^{\rm n} \sim \delta^{3/2}$ with $\delta $ the
overlap between particles, tangential force increment ${\rm d}
F^{\rm t} \sim \delta^{1/2} {\rm d}t$ with ${\rm d}t$ the relative
tangential displacement change, provided $F^{\rm t}\leq \mu F^{\rm
n}$.

\bibitem{palpha}  W. G. Ellenbroek {\em et al.}, 
cond-mat/0604157 (2006).

\bibitem{kostya} K. Shundyak {\em et al.},
cond-mat/0610205 (2006).

\bibitem{footnotefullymob} When fully mobilized contacts are considered
to be {\em fixed} at the Coulomb threshold, gently prepared packings are
found to approach a generalized isostaticity line at small pressures for
any $\mu$ \cite{kostya}, while less gently prepared packings will have
less fully mobilized contacts \cite{silbert,Kasahara}.




\end{thebibliography}
\end{document}